\documentstyle[preprint,aps,floats,tighten]{revtex}
\input{psfig.tex}
\begin{document}
\draft
%\twocolumn[\hsize\textwidth\columnwidth\hsize\csname @twocolumnfalse\endcsname
\preprint{\vbox{\hbox{CU-TP-741} 
                \hbox{CAL-598}
                \hbox{astro-ph/9602150}
}}

\title{Matter/Microwave Correlations in an Open Universe}

\author{Marc Kamionkowski\footnote{kamion@phys.columbia.edu}}
\address{Department of Physics, Columbia University, 538 West
120th St., New York, New York~~10027}
%\date{March 1996}
\maketitle

\begin{abstract}
In an intriguing recent paper, Crittenden and Turok proposed
cross-correlating the cosmic microwave background (CMB) with
tracers of the matter density to probe the existence of a
cosmological constant. Here I
emphasize that a similar cross-correlation arises in an open
Universe and, depending on the redshift distribution of the
tracer population and the matter density, may be comparable to
or stronger than that
in a flat cosmological-constant Universe with the same matter
density. The two cases can be distinguished through
cross-correlation with tracer populations with different
redshift distributions.
\end{abstract}
%]

\vskip 1cm

In an intriguing recent paper, Crittenden and Turok proposed
cross-correlating the
cosmic microwave background (CMB) with tracers of the
matter density to probe the existence of a cosmological constant
$\Lambda$ \cite{crittendenturok}.
In a flat matter-dominated Universe, CMB
anisotropies are produced at (or near) the surface of last scatter
(the Sachs-Wolfe (SW) effect).  If $\Lambda\neq0$, additional anisotropies
are produced as photons pass through potential wells along the
line of sight (the integrated Sachs-Wolfe (ISW)
effect).\footnote{The ISW term discussed here arises in linear
perturbation theory and produces large-angle
CMB anisotropies.  Although the two terms are often used
interchangeably, it should be distinguished from the Rees-Sciama
effect \cite{RS}, in which anisotropies are produced on much
smaller angular scales from nonlinear gravitational collapse of
galaxies or clusters of galaxies.}  Here I re-emphasize that a
similar effect exists 
in an open Universe and point out that, depending on the redshift
distribution of the tracer population, this cross-correlation
can be comparable to or stronger than that in an open Universe
with the same matter density $\Omega_0$.

A cross-correlation between the ISW signal and a tracer of the
mass density at low redshift must be picked out from a noisy
background due to the SW effect.  The contribution to the
signal-to-noise ratio from the $\ell$th multipole moment
therefore depends on the ratio $C_\ell^{\rm ISW}/C_\ell^{\rm tot}$
\cite{crittendenturok}, where $C_\ell^{\rm ISW}$
and $C_\ell^{\rm SW}$ are the ISW and
SW contributions to the $\ell$th  multipole moment.  Although the
precise value of this ratio may depend slightly on the
large-scale power spectrum, it can can be approximated roughly
by \cite{ks,kofman}
\begin{equation}
     {C_\ell^{\rm ISW} \over C_\ell^{\rm SW}} \simeq
     {g(\Omega_0) \over \ell} \equiv {36 \pi \over \ell}
     \int_0^\infty \, \left({dF \over d\eta}\right)^2
     (\eta_0-\eta) d\eta,
\label{onlyequation}
\end{equation}
where $F(\eta)\propto(H/a)\int(da/a)(Ha/a_0)^{-3}$ (normalized
to $F(0)=1$) is the growth factor for gravitational-potential
perturbations as a function of conformal time $\eta$.  Here, $a$
and $H$ are the scale factor and Hubble parameter of the
Universe.

\begin{figure}[htbp]
\centerline{\psfig{file=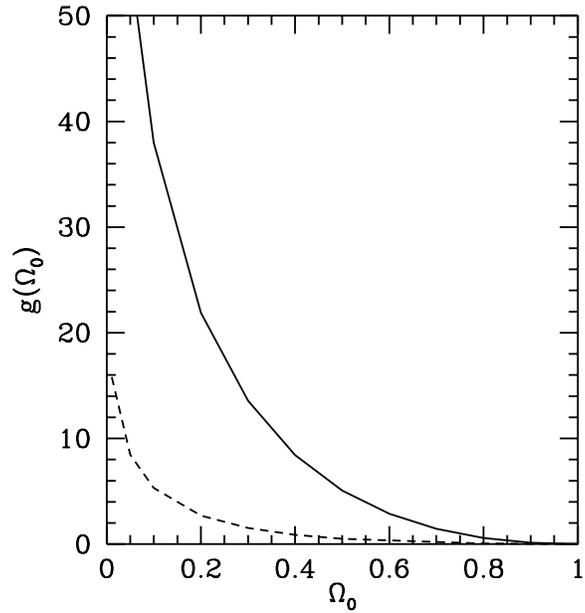,width=3.3in}}
\medskip
\caption{
	The function $g(\Omega_0)$ in an open Universe (solid
	curve) and in a flat $\Lambda$ Universe (dashed
	curve).
}
\label{gplot}
\end{figure}

Figure \ref{gplot} illustrates that $g(\Omega_0)$
is much larger in an open Universe than in a flat $\Lambda$
Universe with the same matter density.
A more accurate treatment of the ISW effect in an open
Universe confirms qualitatively the results presented here
\cite{ks}.  For example, for $\Omega_0=0.3$, the quadrupole is
due almost entirely to the ISW effect (e.g., see Fig. 7 in
Ref.~\cite{ks}), whereas in a flat $\Lambda$ Universe, it
contributes only a fraction of the SW term \cite{kofman}.

\begin{figure}[htbp]
\centerline{\psfig{file=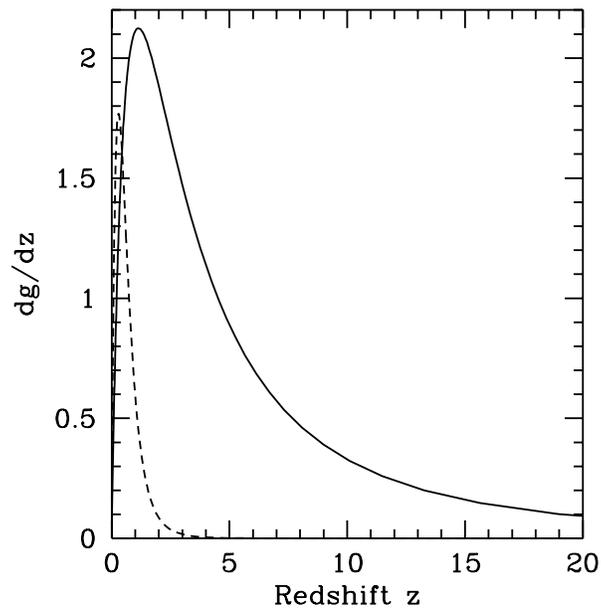,width=3.3in}}
\medskip
\caption{
        The function $dg/dz$ for $\Omega_0=0.3$ in
	an open Universe (solid curve) and in a flat $\Lambda$
	Universe (dashed curve).
}
\label{dgdzplot}
\end{figure}

Realistically, however, the signal will be due to the ISW contribution
from redshifts probed by the tracer population, and the noise will be
due to the SW effect and the ISW contribution from
larger redshifts.  If one has a survey which traces the mass
distribution out to a redshift $z_s$, then to a first
approximation, Eq.~(\ref{onlyequation}) should be replaced by
\begin{equation}
     {C_\ell^{\rm late-ISW} \over C_\ell^{\rm SW} + C_\ell^{\rm
     early-ISW}} \simeq {\int_0^{z_s}\,(dg/dz)\,dz \over \ell +
     \int_{z_s}^{z_{ls}}\, (dg/dz)\,dz},
\label{secondequation}
\end{equation}
where $dg/dz$ the differential contribution to $g(\Omega_0)$ as a
function of redshift $z$ shown in Fig. \ref{dgdzplot}, and
$z_{ls}\simeq1100$ is the redshift of the surface of last
scatter.  For example, suppose we cross-correlate the COBE map
of the CMB with the x-ray background, which probes redshifts up
to $z_s \simeq2$.  Then for $\Omega_0=0.5$,
Eq.~\ref{secondequation} falls in the range 0.11--0.39 for an
open Universe and 0.03--0.19 for a cosmological-constant Universe
for values of $\ell\simeq 2-15$ probed by COBE.  This simple
estimate suggests that the cross-correlation of the CMB with the
x-ray background in an open Universe with $\Omega_0=0.5$ should
at least comparable to that in a cosmological-constant Universe
with the same matter density.  For larger values of $\Omega_0$, the
signal-to-noise becomes larger in an open Universe relative to
its value in a cosmological-constant Universe.  If a tracer
population which extends to a redshift $z_s\simeq5$ can be found, the
signal-to-noise in a cosmological-constant Universe remains
unaffected, but it will increase by more than a factor of two in
an open Universe. 

Of course, more detailed numerical calculations, which take into
account realistic redshift distributions as well as the angular
resolutions and sky coverages of the CMB and tracer surveys,
will be needed for comparison with data.  Still, the estimates
provided here combined with the results of
Ref. \cite{crittendenturok} suggest that this cross-correlation
may provide a useful probe of $\Omega_0$ in an open Universe.
Although these calculations were performed assuming primordial
adiabatic perturbations, a similar cross-correlation should
arise in models with primordial isocurvature perturbations.
More work on large-angle anisotropies in topological-defect
models must be done to determine whether this test will be
effective in these scenarios.

Finally, the flat and open models
should be distinguishable if two (or more) tracer populations
with differing redshift distributions can be used.  This might
also be accomplished by varying the flux cutoff of a single
tracer population.

\bigskip

I thank R. Crittenden for useful discussions.  This work was
supported in part by the D.O.E. under contract DEFG02-92-ER
40699 and by NASA under contract NAG5-3091.

\end{document}